\documentclass[printer]{aa}
\usepackage[english]{babel}
\usepackage{txfonts}
\usepackage{natbib}
\usepackage{aas_macros}
\usepackage{graphicx}
\def\igr {\emph{INTEGRAL} }
\def\sax {\emph{BeppoSAX} }
\def\xmm {\emph{XMM-Newton} }
\def\src {SGR\,1900+14 }
\def\sgr {SGRs }
\def\flux {\mbox{erg cm$^{-2}$ s$^{-1}$}}
\def\lum {\mbox{erg s$^{-1}$}}

\begin{document}
\title{Five years of \src observations with \emph{BeppoSAX}}
\author{P. Esposito\inst{1,2}
    \and S. Mereghetti\inst{2}
    \and A. Tiengo\inst{2}
    \and L. Sidoli\inst{2}
    \and M. Feroci\inst{3}
    \and P. Woods\inst{4,5}}
\institute{Universit\`a degli Studi di Pavia, Dipartimento di Fisica Nucleare e Teorica and INFN-Pavia, via Bassi 6, I-27100 Pavia, Italy
    \and INAF - Istituto di Astrofisica Spaziale e Fisica Cosmica Milano, via Bassini 15, I-20133 Milan, Italy
    \and INAF - Istituto di Astrofisica Spaziale e Fisica Cosmica Roma, via Fosso del Cavaliere 100, I-00133 Rome, Italy
    \and Dynetics, Inc., 1000 Explorer Boulevard, Huntsville, AL 35806
    \and National Space Science and Technology Center, 320 Sparkman Drive, Huntsville, AL 35805}
\offprints{P.~Esposito,\\
    \email{paoloesp@iasf-milano.inaf.it}}
\date{Received 2006 May 3 / Accepted 2006 August 31}
\abstract{ We present a systematic analysis of all the \sax data of
the soft gamma-ray repeater SGR\,1900+14: these observations allowed
us to study the long term properties of the source quiescent
emission. In the observation carried out before the 1998 giant flare
the spectrum in the 0.8--10 keV energy range was harder and
there was evidence for a 20--150 keV emission, possibly associated
with SGR\,1900+14. This possible hard tail, if compared with the
recent \igr detection of SGR\,1900+14, has a harder spectrum
(power-law photon index $\sim$1.6 versus $\sim$3) and a
\mbox{20--100 keV} flux $\sim$4 times larger. In the last \sax
observation (April 2002), while the source was entering the long
quiescent period that lasted until 2006, the \mbox{2--10 keV} flux
was $\sim$25\% below the historical level. We also studied in detail
the spectral evolution during the 2001 flare afterglow. This was
characterized by a softening that can be interpreted in terms of a
cooling blackbody-like component.
    }
\titlerunning{X-ray spectra of \src }
\authorrunning{Esposito et al.}
\maketitle
\section{Introduction}
Soft Gamma-ray Repeaters (SGRs) are a small group of high-energy sources, originally discovered through the emission of their characteristic short $\gamma$-rays bursts. Only four confirmed \sgr are known, plus two candidates. SGR bursts have typical duration of the order of 0.1 s, peak luminosity in the \mbox{$10^{39}$\textbf{--}$10^{42}$ erg s$^{-1}$} range, and are emitted during ``active'' periods that can last from weeks to months. Exceptionally large outbursts are also observed in SGRs. These rare events have properties clearly different from those of the usual short \mbox{$\gamma$-ray} bursts. Based on their intensities, they can be classified either as ``giant'' flares, with total released energy up to \mbox{$10^{47}$ erg}, or ``intermediate'' flares, with total energy  smaller by orders of magnitude ($10^{41}$--$10^{43}$ erg). In the classical X-ray range \mbox{($\sim$1--10 keV)} SGRs are relatively steady sources with luminosity in the $10^{35}$--$10^{36}$ \lum range (although fainter states have also been observed, see \citealt{kouveliotou03} and \citealt{mereghetti06}) and showing periodic pulsations with periods of several seconds and secular spin-down of \mbox{$\sim$$10^{-11}$--$10^{-10}$ s s$^{-1}$}.\\
\indent It is generally thought that SGRs, as well as a group of similar pulsars known as Anomalous X-ray Pulsars (AXPs), are magnetars, i.e. isolated neutron stars with strong magnetic fields (see \citealt{woods04} for a review of this class of objects). In the magnetar model both the persistent X-ray emission and the bursts are powered by magnetic energy \citep{duncan92,thompson95,thompson96}. If the secular spin-down observed in \sgr is attributed to dipole radiation losses, as in ordinary radio pulsars, magnetic fields of the order of $10^{14}$--$10^{15}$ G are inferred.\\
\indent In this paper we focus on SGR\,1900+14, reporting all the observations of this source carried out with the \sax satellite. Although some of these data have been already published \citep{woods99b,woods01,feroci03}, we reanalyzed all the data sets following the same procedure, in order to compare them in a consistent way. In fact these observations, spanning five years and covering different states of bursting\,/\,flaring activity, give the possibility to investigate the long term  spectral and flux variability of the source with a homogeneous data set.\\
\indent In Section \ref{review} we briefly review some results on SGR 1900+14, in the context of the activity history of the source. The spectral  and timing analysis are reported in Sections 3 and 4, where we focus on the long term changes in the 1--10 keV emission properties. In Section 5 we report evidence for the detection of SGR 1900+14 in the \mbox{20-150 keV} band during one of the \sax observations. In Section 6 we concentrate on the spectral variability on short time-scales  following the April 2001 intermediate flare.
\section{SGR 1900+14: activity episodes and \sax observations}\label{review}
This SGR was  discovered  in 1979  when a  few bursts were recorded with the \emph{Venera 11} and \emph{Venera 12} satellites \citep{mazets79}. No other bursts were detected until thirteen years later, when four more events were seen with the  BATSE instrument on the  \emph{Compton-GRO} satellite in 1992 \citep{kouveliotou93}. The X-ray counterpart, discovered with  \emph{ROSAT} \citep{vasisht94}, was observed a first time with \sax \citep{woods99b}. The periodic pulsations in the X-ray counterpart (period of $\sim$5.2 s)  were discovered with the  \emph{ASCA} satellite during an observation in April 1998 \citep{hurley99}, which took place just three weeks before the burst reactivation of the SGR \citep{hur99}. Subsequent observations  with the \emph{Rossi-XTE} satellite confirmed the pulsations and established that the source was spinning down rapidly, with a period derivative of $\sim$$10^{-11}$ \mbox{s s$^{-1}$} \citep{kouveliotou99}.\\
\indent The peak of the bursting activity  for \src was reached on 1998 August 27, when a giant flare was recorded by numerous instruments. This flare started with a short (\mbox{$\sim$0.07 s}) soft spike (often referred to as the ``precursor''), followed by a much brighter hard pulse (duration $\sim$1 s) that reached at least $\sim$$10^{45}$ \mbox{erg s$^{-1}$} and a soft $\gamma$-ray tail modulated at 5.2 s \citep{hurley99gf,maz99b,feroci01gf}. The oscillating tail decayed quasi-exponentially over the next $\sim$6 minutes \citep{feroci01gf}. Integrating over the entire flare assuming isotropic emission, at least $10^{44}$ erg were released in hard X-rays above 15 keV  \citep{maz99b}. Another bright burst was detected on August 29 \citep{ibrahim01}, scaled down by a factor of $\sim$100 in peak luminosity and duration, compared to the August 27 flare. The second \sax observation was done less than one month after these events when the source was still active and showed an enhanced X-ray emission \citep{woods99b}.\\
\indent After almost two years of quiescence, during which two \sax observations were carried out \citep{woods01}, \src emitted an intermediate flare on 2001 April 18 \citep{guidorzi01}. This event, which prompted the two following \sax observations \citep{feroci03,feroci04,woods03}, had a duration of $\sim$40 s and a total fluence of $1.6\times 10^{-4}$ \mbox{erg cm$^{-2}$}. Another bright flare, but of comparatively smaller fluence ($\sim$$9\times 10^{-6}$ \mbox{erg cm$^{-2}$}), occurred after 10 days \citep{lenters03}. The last bursts reported from SGR\,1900+14, before its recent reactivation \citep{palmer06,golenetskii06}  occurred in November 2002 \citep{hurley02}.\\
\indent All the \sax observations of \src are listed\footnote{Observations G and H are listed for completeness, but are of scarce utility due to their short integration time and presence of contaminating sources in the PDS instrument; they will not be discussed further.} in Table \ref{obs}. In each observation the SGR was aligned with the optical axis of the instruments. In  summary: three observations were triggered by the occurrence of flares (B, E and F), and took place while the source was still active, as testified by the detection of bursts in the \sax data, while all  the other observations can be considered as representative of the source quiescent state emission.
\section{Spectral Analysis}
\begin{figure}[h!]
\resizebox{\hsize}{!}{\includegraphics[angle=90]{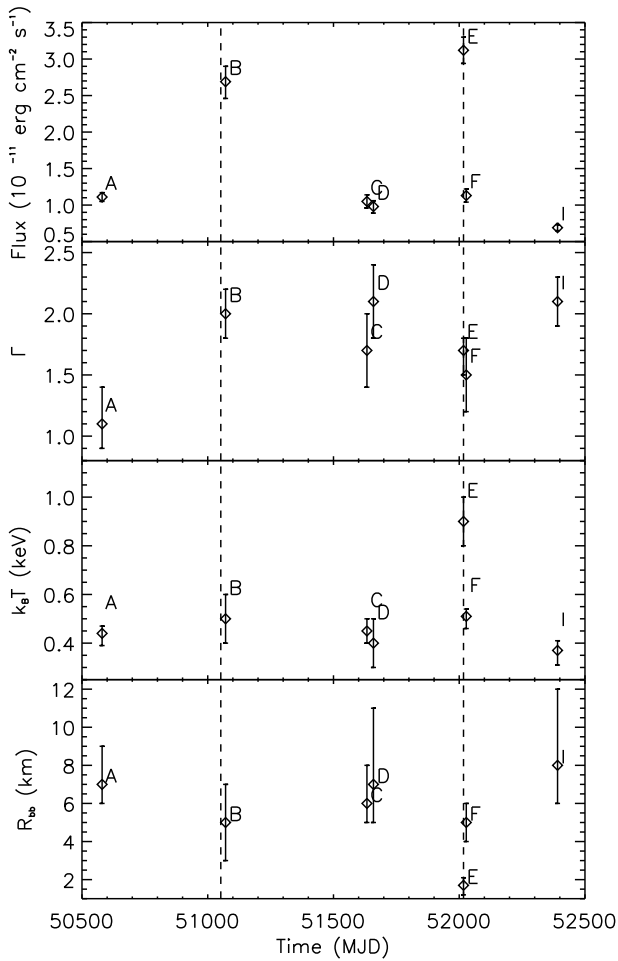}}
\caption{Long term evolution of the 2--10 keV unabsorbed flux and of the spectral parameters of SGR\,1900+14 (assuming for the absorption the value of $2.6\times10^{22}$ cm$^{-2}$). The vertical dashed lines indicate the time of the 1998 August 27 giant flare and of the 2001 April 18 intermediate flare. The error bars are at the 90\% confidence level.}
\label{1900_lt_lc}
\end{figure}

The results presented in this section were obtained with the Low Energy Concentrator Spectrometer (LECS) and the Medium Energy Concentrator Spectrometer (MECS) instruments \citep{parmar97,boella97mecs}. Both are imaging detectors operating in the 0.1--10 keV and 1.8--10 keV energy ranges respectively.\\
\indent We used source extraction regions with radii of $4\arcmin$ and $8\arcmin$ for the MECS and the LECS, respectively. Because of the low Galactic latitude of SGR\,1900+14, in order to properly account for the presence of the diffuse emission from the Galactic Ridge, concentric rings of \mbox{6\arcmin.4--9\arcmin.6} and \mbox{$9\arcmin$--$13\arcmin$} were chosen from each pointing for background subtraction with the MECS and the LECS, respectively. The bursts in observations B, E and F were excluded from the analysis\footnote{The spectral results for the bursts detected in observation E are reported in \citet{feroci04}.}. This was done by extracting light curves with a bin size of 1 s and applying intensity filters. All the spectra were rebinned to achieve at least 30 counts in each spectral channel and to oversample by a factor 3 the instrumental energy resolution. The fits were performed simultaneously, over the energy ranges 1.8--10 keV (MECS) and 0.8--4.0 keV (LECS), and including a constant factor to account for normalization uncertainties between the instruments (this factor was constrained to be within its usual range\footnote{See the Cookbook for \emph{BeppoSAX} NFI Spectral Analysis, \mbox{\texttt{http://www.asdc.asi.it/bepposax/software/cookbook/}}}). Spectral analysis has been performed with the \textrm{XSPEC v.11.3.2} software package \citep{arnaud96}.\\
\indent In some observations a fit with an absorbed power-law yields
unacceptable $\chi^{2}$ values, therefore we explored a power-law
plus blackbody model which gave good fits for all the data sets.
Since there is no obvious physical reason for the absorption to
change, at least while the source is in quiescence, we fitted all
the data sets also with a common value for the $N_{\rm H}$. The
value of \mbox{$2.6\times10^{22}$ cm$^{-2}$} has been derived
fitting simultaneously the spectra of the observations performed
while the source was in quiescence. The blackbody temperature
($\sim$0.4 keV) and emitting area\footnote{We assume for \src a
distance of 15 kpc \citep{vrba00}.} (\mbox{R$\sim$6-7 km}) do not
vary much, except during observation E. This observation was
performed during the afterglow of the 2001 April 18 flare, and shows
clear evidence for spectral variations within the observation (see
Section \ref{obsE}). In Figure \ref{1900_lt_lc} we have plotted the
long term evolution of the flux and spectral parameters obtained in
the power-law plus blackbody fits and all the best fit parameters
are reported in Table \ref{fits}. They are consistent with  those
obtained by \citealt{woods99b} (for observations A and B),
\citealt{woods01} (C and D), and \citealt{feroci03} (E and F).
\\

\begin{figure}[h]
\resizebox{\hsize}{!}{\includegraphics[angle=270]{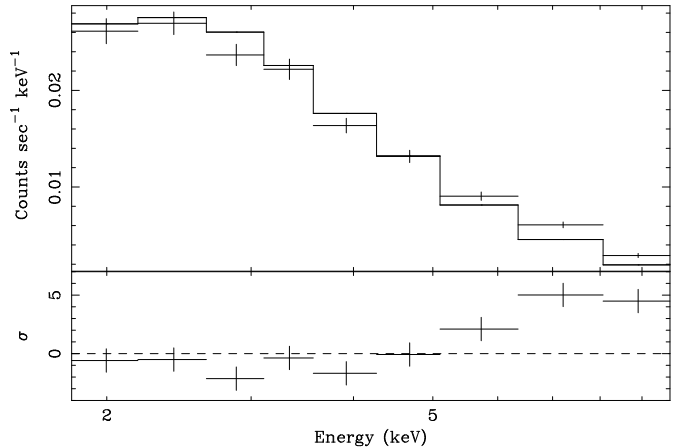}}
\caption{{\it BeppoSAX}\,/\,MECS spectrum of observations A and
residuals with respect to the simultaneous fit of the spectra of
observation A, C, D, F, and I with an absorbed power-law plus
blackbody model with only an overall normalization factor left free
to vary. The data have been rebinned graphically to emphasize the
trend in the spectral residuals.} \label{1900res}
\end{figure}

\indent The observations in which SGR 1900+14 had the highest X-ray
flux are those following the two flares (B and E). The flux in the
last observation (I), whose analysis is reported here for the first
time, is instead $\sim$25\% lower than in the other quiescent
observations. The fading  is also confirmed by a simple comparison
of the MECS count rates of observations I and D, which differ at $>
$10$\,\sigma$ level. During observation I the transient source
XTE\,J1908+94 \citep{intzand02}, located $24\arcmin$ from the SGR
(i.e. just inside the MECS field of view), was in a high state.
Therefore we carefully checked our flux estimate for SGR 1900+14 by
exploring different background and source extraction regions. Our
conclusion is that the observed decrease in the flux is real.\\
\indent Figure \ref{1900_lt_lc} also shows that the power-law
component during observation A was slightly harder than in
all the following quiescent state observations, performed after the
1998 August 27 giant flare.
In order to compare the hardness of the overall spectra of
the quiescent observations, we have simultaneously fit them with the
same parameters (introducing a normalization factor to account for
the flux change) and we note that the spectra C, D, F, and I give an
acceptable fit, while the addition of spectrum A makes the
simultaneous fit unacceptable, due to the high energy excess shown
in figure \ref{1900res}. This means that the pre-flare spectrum was
significantly harder than the average quiescent spectrum of SGR
1900+14 measured by {\it BeppoSAX} after the giant flare.

\section{Timing analysis}
For the timing analysis we first corrected the time of arrival of
the MECS events to the solar system barycenter, and then used
standard folding techniques to measure the source spin period. For
observation I we find a period of \mbox{$5.18019\pm0.00002$ s}, and
for all the other observations our values (reported in Table
\ref{obs} and in Figure \ref{1900_lt_lc}) are in agreement with
those of \citet{woods99b,woods01} and \cite{feroci02}. In Figure
\ref{pp} we show the background-subtracted phase-folded profiles and
the pulsed fractions for the seven data sets. We derived the pulsed
fractions and the relative errors fitting the pulse profiles with a
constant plus one or two (for observation A) sinusoidal functions
and computing the ratio between the sin amplitude and the constant.
\begin{figure}[h!]
\resizebox{\hsize}{!}{\includegraphics[angle=90]{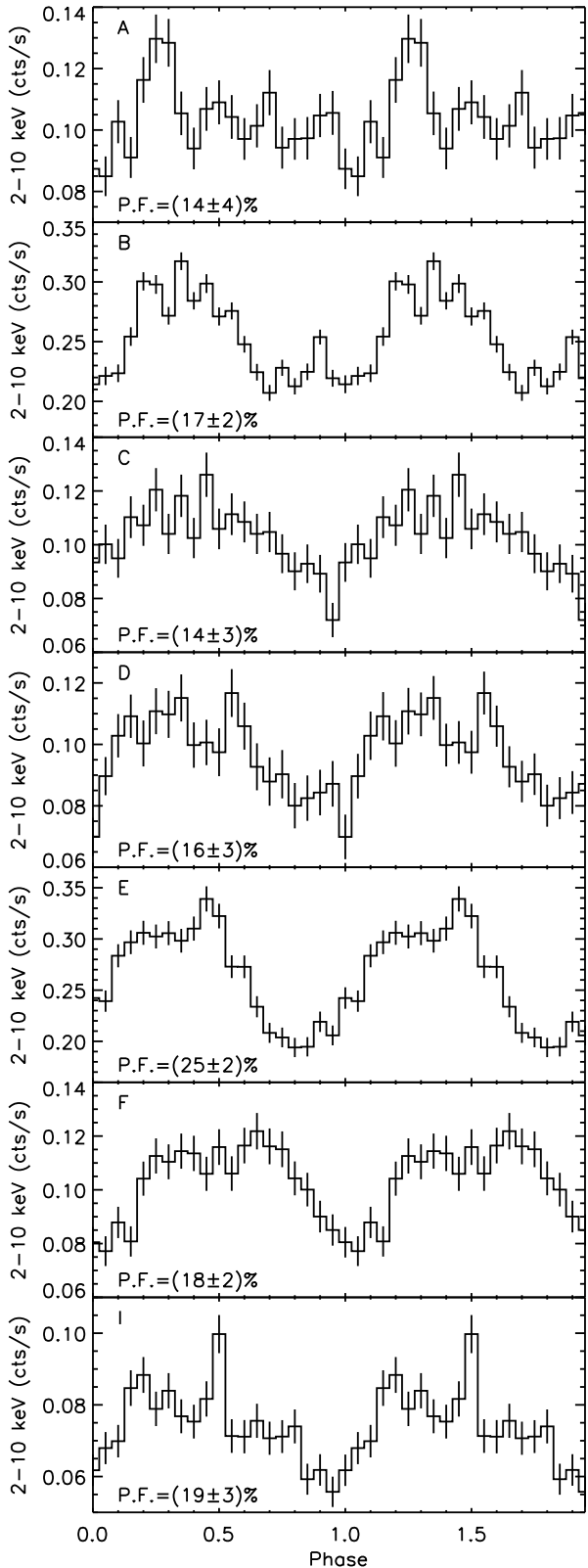}}
\caption{MECS pulse profiles (not phase-connected) and pulsed fraction of \src in the seven observations (as indicated in the panels).}
\label{pp}
\end{figure}
Although  \src has been extensively monitored with the \emph{RXTE} satellite and there are detailed studies of its light curve and pulsed flux evolution (see e.g. \citealt{woods99d} and \citealt{gogus02}), the results presented here, being obtained with an imaging instrument, have the advantage of providing absolute flux and pulsed fraction measurements.\\
\indent We note that the changes in the spectrum and in the pulse profile after the giant flare were not accompanied by significant variations in the pulsed fractions. The only significant change has been measured during observation E, when the pulsed fraction was higher ($\sim$25\%) then the average value of $\sim$17\% (this enhancement, related to the afterglow emission, has been discussed in \citealt{feroci03}). In contrast the pulse shape of SGR\,1806$-$20, the only other SGR observed before and after a giant flare, was only slightly different after the event, and its pulsed fraction remained small ($\sim$4\%, \citealt{tiengo05,rtm05}) until two months after the flare and then increased to the pre-flare value ($\sim$10\%, \citealt{rtm05,woods06}).\\
\section{Hard X-ray detection}
The PDS instrument (Phoswich Detection System, \citealt{frontera97}) extended the spectral and timing capabilities of \sax to the 15--300 keV band. This non-imaging spectrometer had a field of view of $1.3\degr$ (FWHM) and the background subtraction was done with a rocking system, which switched between the source and two background regions offset by $3.5\degr$ every 96 s.\\
\indent In all the PDS exposures listed in Table \ref{obs} we detected a significant hard X-ray emission. However three transient X-ray sources, the pulsars 4U\,1907+97 \citep{giacconi71,liu00} and XTE\,J1906+09 \citep{marsden981906}, and the black hole candidate XTE\,J1908+94 \citep{intzand02}, are located at a small angular distance from \src ($47\arcmin$, $33\arcmin$ and $24\arcmin$ respectively). When in high state, they can reach fluxes above \mbox{$\sim$$10^{-9}$ \flux} in the \mbox{20--100 keV} range, preventing a sensitive search for a (presumably dimmer) emission from SGR\,1900+14. XTE\,J1908+94 during a bright state is clearly identifiable in the MECS and LECS, since it lies within the field of view of these imaging instruments. This was the case of observation I, performed shortly after the discovery of that source \citep{woods02IAUC}. The presence of the other two sources has been identified from the detection in the PDS of periodicity at their known pulse periods (\mbox{$\sim$89 s} for XTE\,J1906+09 and \mbox{$\sim$440 s} for 4U\,1907+97) in all the PDS data sets except in the first one. Therefore only for the 1997 observation there is no evidence of contamination from one of these three sources. Given that \src lies at a low Galactic latitude (b=0.77\degr), we might worry that the flux observed in the PDS during observation A could result from diffuse emission from the Galactic Ridge. Since this emission is constant in time, the lowest count rate observed in later observations (see Table \ref{pds}) allows us to set an upper limit to its contribution in observation A. This upper limit is of $\sim$60\% of the detected flux in the \mbox{20--50 keV} band and of $\sim$10\% in the \mbox{50--150 keV} band. Although we cannot rule out the possibility of contamination from unknown transient sources, we conclude that the flux measured in observation A up to $\sim$150 keV is very likely due to SGR\,1900+14.\\
\indent We extracted the PDS background subtracted spectrum and
using the most recent response matrix, we fitted the logarithmically
rebinned PDS spectrum in the range \mbox{15--150 keV}. With a simple
power-law model we obtained a photon index $\Gamma=1.6\pm0.3$ and a
20--100 keV flux of \mbox{$(6\pm1)\times10^{-11}$ \flux} with a
$\chi^2_r$ value of 0.98 for 40 d.o.f. . We also fitted the PDS
spectrum simultaneously  with the LECS and MECS spectra, using a
blackbody plus power-law model. We included a factor to account for
normalisation uncertainties between the low-energy instruments and
the PDS. This factor assumed the value of 0.90 (the range of
acceptable values is \mbox{0.77--0.95}). The resulting best fit
parameters (photon index $\Gamma=1.04\pm0.08$, blackbody temperature
$k_BT=0.50\pm0.06$, radius $R_{bb}=5\pm2$ km, and absorption $N_{\rm
H}=(1.8\pm0.5)\times10^{22}$ cm$^{-2}$) are consistent with an
extrapolation of the power-law component measured at lower energy
(Figure \ref{sed}.a). We also checked a broken power-law plus
blackbody model and, although the improvement in the goodness of the
fit, as measured by the F-test statistic, is marginal, we obtained a
slightly lower $\chi^2$ value (1.11 for 135 d.o.f. instead of 1.17
for 137 d.o.f.) with a photon index of $\sim$0.7 up to $\sim$25 keV
and of $\sim$1.7 above, and with a similar blackbody component
(Figure \ref{sed}.b). Motivated by the structured residuals from
$\sim$15 keV to $\sim$35 keV, where, as discussed above, some
contamination from the Galactic diffuse emission cannot be excluded,
we performed also a fit using the PDS data only above 35 keV. The
resulting parameters are photon index $\Gamma=1.15\pm0.10$,
blackbody temperature $k_BT\simeq0.5$ keV and $R_{bb}\simeq5$ km,
with a $\chi^2_r$ value of 1.08 for 123 d.o.f. . The 20--100 keV
flux derived from all the fits is \mbox{$(7\pm1)\times10^{-11}$
\flux}.
\begin{figure}[h]
\resizebox{\hsize}{!}{\includegraphics[angle=270]{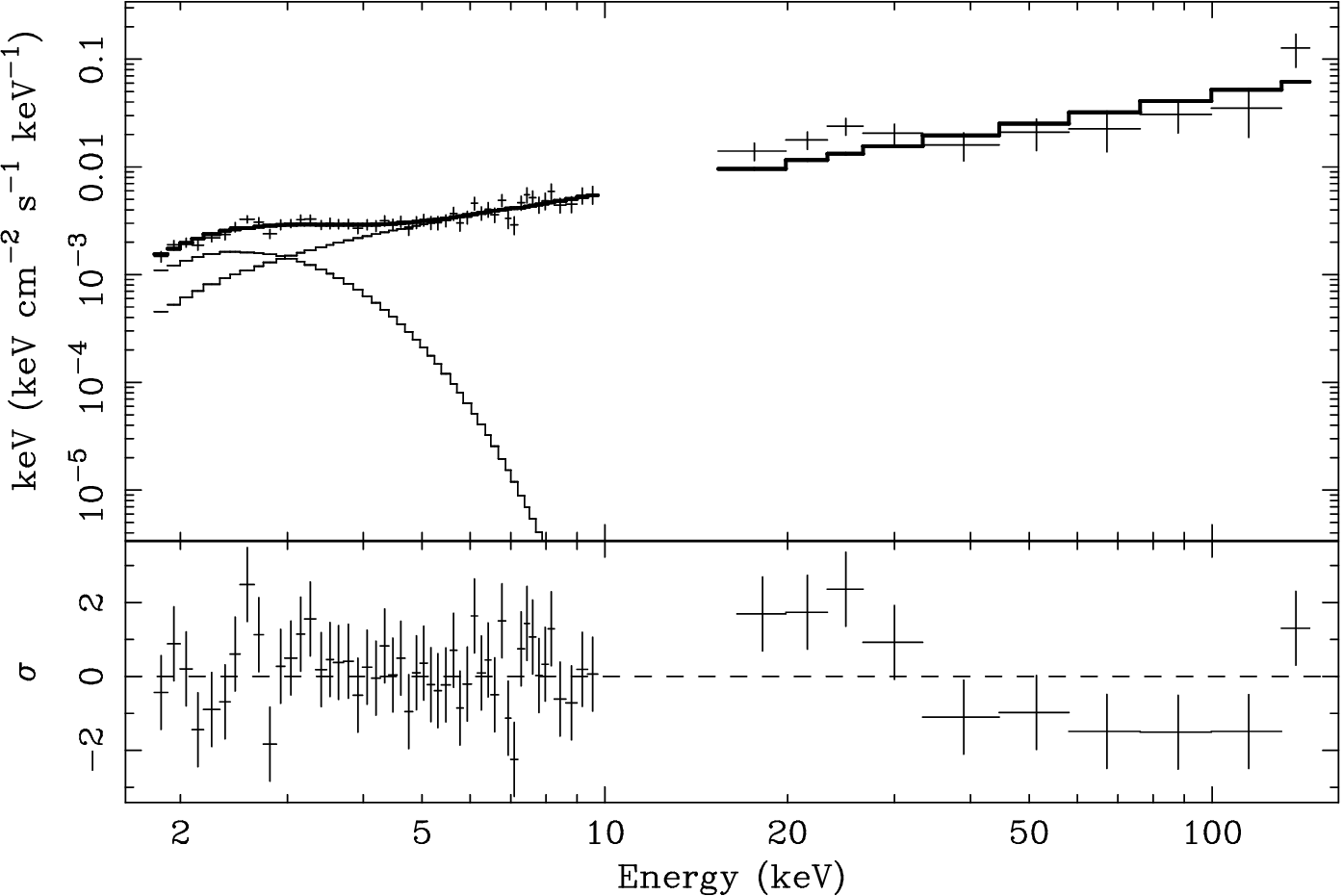}}
\setcounter{figure}{2}
\caption{\textbf{{\small a.}}}

\vspace{0.1cm}

\resizebox{\hsize}{!}{\includegraphics[angle=270]{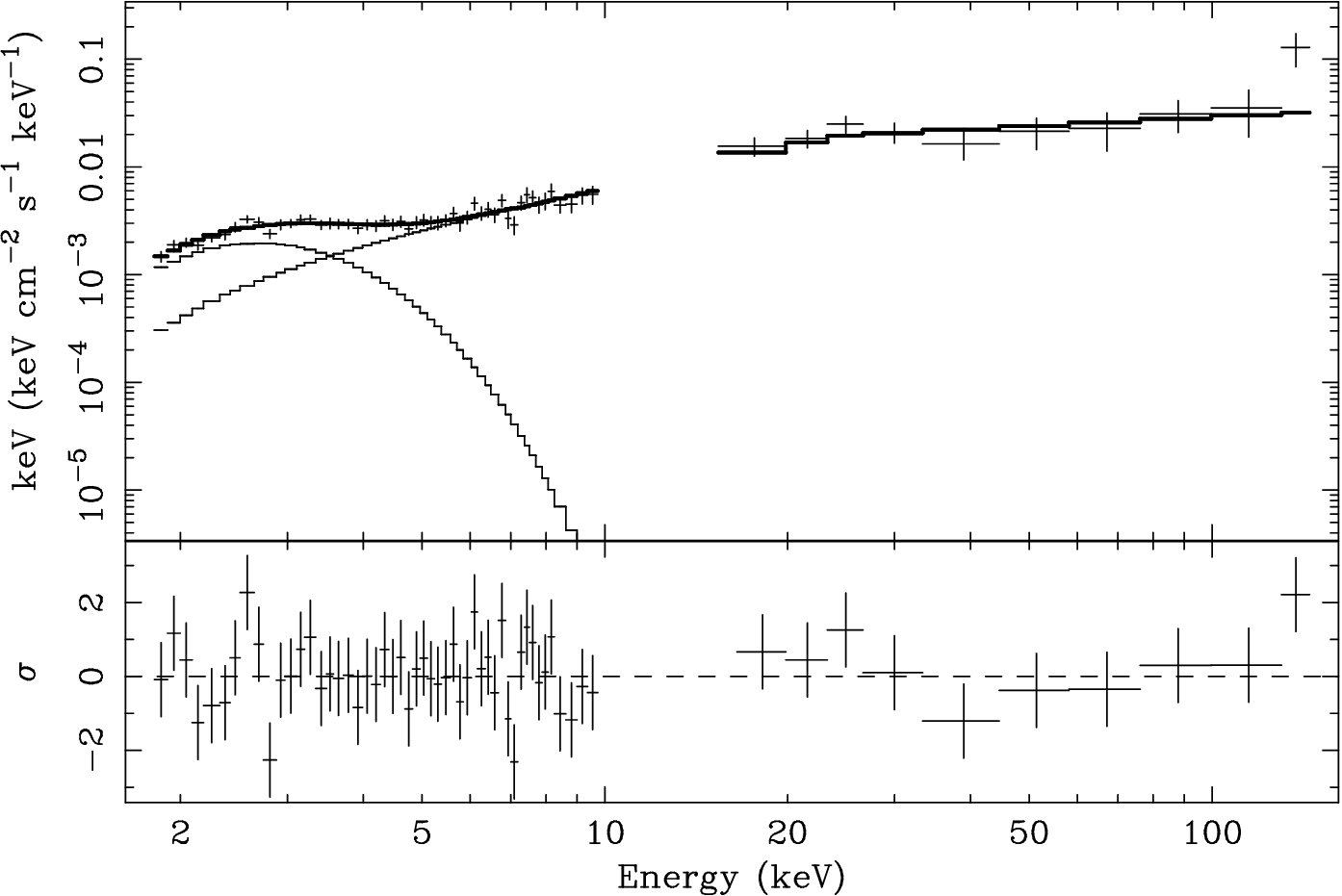}}
\setcounter{figure}{2}
\caption{\textbf{{\small b.}} Broad band spectrum and residuals from the data of the observation A fitted with a power-law plus blackbody model (\emph{a}) and with a broken power-law plus blackbody model (\emph{b}). The data points are from the MECS and PDS instruments and the thick line represents the total model, while the thin lines represent its absorbed power-law and blackbody components.}
\label{sed}
\end{figure}\\
\indent In order to search for pulsations in the hard X-ray range we folded the PDS data at the period of \mbox{5.15719 s} measured with the MECS, but no significant periodic signal was detected. The 3\,$\sigma$ upper limit on the source pulsed fraction derived by a sinusoidal fit is $\approx$50\%.\\
\indent Except for observation H, whose high count rate is due to
XTE\,J1908+94\footnote{The source went in outburst in February 2002
and reached its flux peak about two months later
\citep{intzand02}.}, all the post-giant flare observations show a
lower count rate in the \mbox{50--150 keV} band with respect to
observation A. The consistent count rates obtained in every
observations from the two uncorrelated regions used for background
subtractions (see Table \ref{pds}) assure that this decrease does
not result from bright sources in the background pointings.
Moreover, the contamination in this band from the X--ray
pulsars is expected to be negligible in all observations, since
their spectrum in outburst is characterized by a high energy cutoff
at 10--15 keV \citep{wilson02,baykal06}.

This considerations lead us to conclude that \src became less bright
in the \mbox{50--150 keV} band after its giant flare. The
fact that the 20--50 keV count rate during observation C was lower
than in observation A, even though the pulsar XTE\,J1906+09 was
active, might indicate that the flux of \src\ in this softer energy
band had also significantly decreased.
\section{Spectral variability in the afterglow of the 18 April
2001 flare}\label{obsE}
\begin{figure}[h!]
\resizebox{\hsize}{!}{\includegraphics[angle=0]{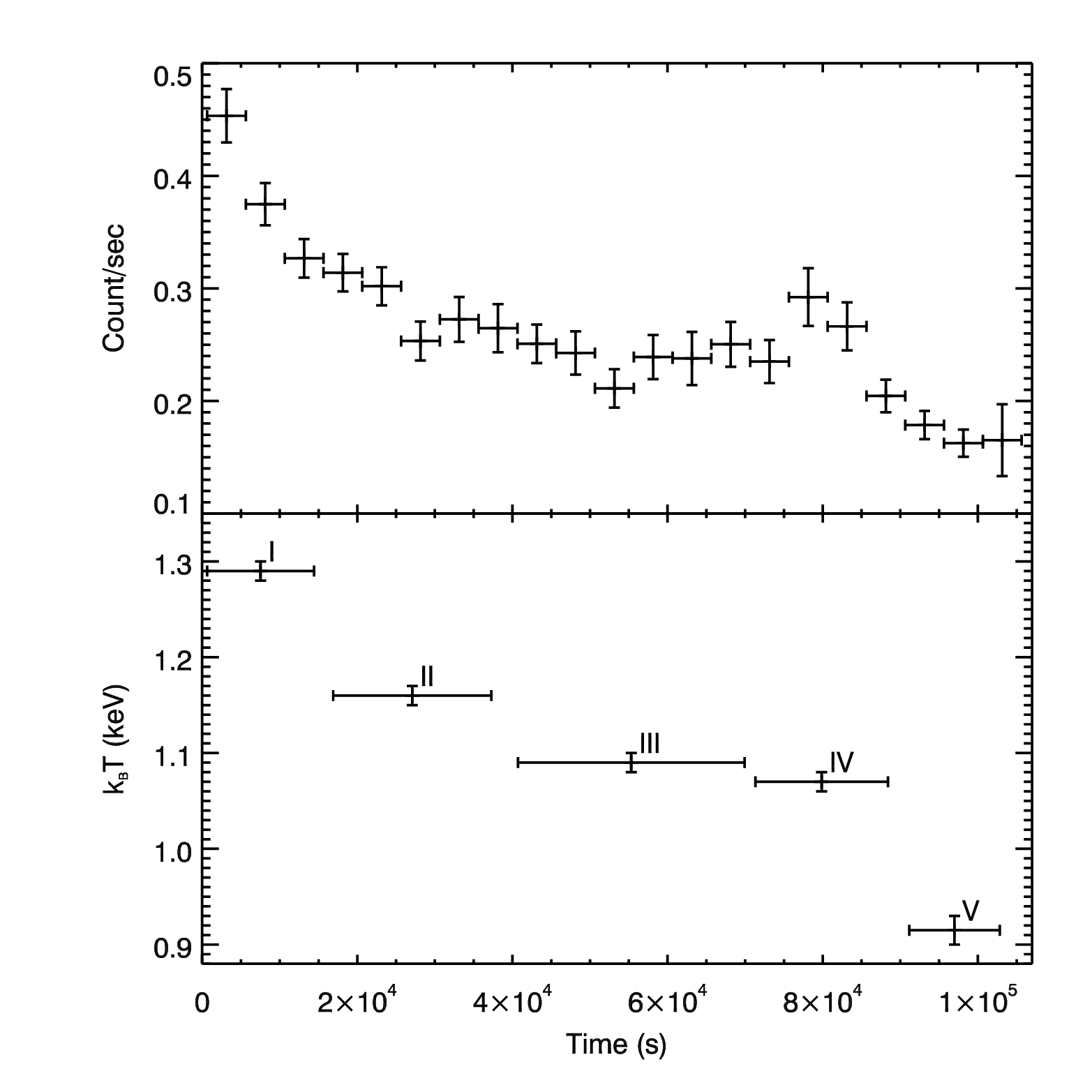}}
\caption{Background subtracted MECS 2--10 keV light curve and blackbody temperature observed on 2001 April 18 about 7.5 hours after the flare. See Table \ref{fitE} for the latter values, obtained from the addition of a new blackbody component with fixed emitting area. The time intervals with bursts have been excluded. Error bars are at 1\,$\sigma$.}
 \label{lc_E}
\end{figure}

Flux and spectral variations as a function of time within the individual observations (except for the bursts) were evident, as mentioned above, only for the data collected $\sim$7.5 hours after the onset of the 2001 April 18 flare (observation E). While evidence for this based only on hardness ratio analysis was reported in \citet{feroci03}, here we present, for the first time, a time resolved spectral analysis of the afterglow lightcurve.\\
\indent The \src light curve for this observation, binned in \mbox{5\,000 s} intervals, is shown in the top panel of Figure \ref{lc_E}. A detailed study of the flux decay, using also data from RXTE and Chandra that filled the time gap between observations E and F, has been reported by \citet{feroci03}. They showed that, after subtracting a constant flux corresponding to the pre-flare quiescent level, the light curve is well described by a power-law with $F\propto t^{-0.9}$, with superimposed a broad ``bump'' (visible at $t$ $\sim$ \mbox{80\,000 s} in Figure \ref{lc_E}).
\\
\indent Following this approach, we assumed that the variable ``afterglow'' emission is present on top of a ``quiescent'' emission that shows only moderate variations on long time-scales. We therefore extracted the source spectra for five different time intervals (our selection is visible in Figure \ref{lc_E}) and fitted them with a model consisting of a power-law plus blackbody with fixed parameters, plus a third variable component to model the afterglow emission. As representative  parameters and normalization of the fixed emission we used values consistent with those seen in the last observations before the flare (C and D), i.e. $\Gamma\simeq2$, \mbox{$k_BT\simeq0.4$ keV} and \mbox{$R_{bb}\simeq7$ km}. We found that the variable component was better described by a blackbody than by a power-law (typical $\chi^2_r$ values of $\sim$1.1 and $\sim$1.7, respectively). The results for the blackbody fits are reported in Table \ref{fitE}. Relatively good fits were also obtained by imposing either a constant temperature or a constant emitting area along the whole observation. The temperatures derived in the latter case (that generally gives lower $\chi^2$ values) are plotted in the bottom panel of Figure \ref{lc_E}. These results show that a cooling  blackbody emission from a region of constant surface could account for both the flux decrease and the spectral softening observed during the afterglow.\\
\indent However we note that, due to the relatively low statistics of the time resolved spectra, other spectral decompositions are consistent with the data. One possibility is for example to use the power-law plus blackbody model adopted for the time integrated emission with only either the power-law or the blackbody parameters free to vary.
\section{Discussion}
Our re-analysis of the \sax data of \src confirms the spectral variability found in this source by \citet{woods99b,woods01}, and \citet{feroci03} on yearly time-scale. Since they found that in some observations an additional blackbody component was required, we were interested in a more thorough assessment of its possibly persistent presence. Such a two-components spectrum is one of the characteristics of the AXPs \citep{mereghetti02} and has also been observed in the other well studied soft gamma-ray repeater \mbox{SGR\,1806--20} \citep{mte05}. Although formally required only in two (possibly three) observations, that component might well be a permanent feature, always present in this source. In fact, except during the aftermath of the April 2001 flare, its temperature ( \mbox{$\sim$0.4 keV}) and emitting area (\mbox{$\sim$6--7 km}) are consistent with all the spectra. If we assume an underlying and nearly steady blackbody, it might be that, as proposed by \citealt{woods99b} and \citealt{kouveliotou01}, this spectral component is visible only in the observations that offer both a low power-law flux and good statistics.\\
\indent The long term spectral variability seems to correlate with the occurrence of the giant and intermediate flares and, in a more complex way, with the ordinary bursting activity. Comparing the only \sax pre-flare observation with the quiescent post-flare ones,  there is evidence for a softening in the spectrum. Also SGR\,1806$-$20 after its 2004 December 27 giant flare displayed a softer spectrum with respect to the 2004 levels \citep{rtm05}. This is qualitatively consistent with the magnetar scenario, in which the spectral hardening is linked to the increasing torque of the twisted magnetosphere, that finally drives the SGR to a giant flare \citep{tlk02,mte05}. Then, after the flare, the source magnetosphere is foreseen to relax into a less twisted configuration, with a softer spectrum.\\
\indent The most recent \sax observation of \src (Observation I, April 2002), shows a small but statistically significant fading compared to the preceding observations. A long term monotonic decrease of the X--ray emission has been observed in SGR\,1627$-$41 \citep{kouveliotou03,mereghetti06} from 1998 to 2004. During this period no bursts were recorded from SGR\,1627$-$41, and its fading has been interpreted as due to the cooling of the neutron star surface after the heating occurred when the source was active in 1998. \src was still moderately active during 2002 \citep{hurley02}, but then no bursts were observed for several years. The smaller luminosity in the last \sax observation might thus correspond to the initial part of a cooling and fading phase, at least qualitatively similar to that observed in SGR\,1627$-$41, but now interrupted by the  recent (March 2006) reactivation  \citep{palmer06,golenetskii06}.\\
\indent During the afterglow of the 2001 April 18 flare, \citet{feroci03} found a flux decrease and a spectral softening. Our re-analysis shows that the variable spectral component can be well modeled as an additional blackbody emitted from a smaller and hotter (but rapidly cooling) region of the neutron star surface. Successful attempts to explain observations of afterglow flux decays in magnetars by means of a cooling thermal component are described in \citet{ibrahim01}, \citet{lenters03}, and \citet{woods04axp}; all these works point out evidence of cooling hot spots on the surface of the neutron star exposed to a fireball.  However we note that the occurrence of the bump in the light curve of the afterglow is an anomaly in the picture of the cooling of a thermal emission, since it requires a re-injection of energy; we refer to \citealt{feroci03} for an extensive discussion of this issue.\\
\indent Evidence for persistent emission above 20 keV for \src has
recently been obtained with \igr observations \citep{gotz06}. We
found that a hard tail was visible also in the 1997 \sax PDS data.
If this emission is indeed due to SGR\,1900+14, our 50 ks long
observation indicates significant differences with respect to the
average properties obtained with \igr, based on the sum of many
observations performed discontinuously  from March 2003 to June
2004. The PDS \mbox{20--100 keV} flux is  $\sim$4 times
larger\footnote{The uncertainty in the relative calibration of the
two satellite in the energy band considered here is of $\approx$10\%
\citep{kirsch05}.} and the spectrum is harder (photon index
$\sim$1.1) than that measured with \igr (photon index $\sim$3). Even
considering our fit based only on the PDS instrument, the difference
in the hard X-ray spectral index is significant (photon index
$\sim$1.6 versus $\sim$3).
Another interesting indication from the PDS data is a decrease of the \mbox{50--150 keV} flux of \src after the giant flare: it is possible that the hard X-ray flux decrease and softening in \src was  a consequence of the 1998 August 27 giant flare.\\
\indent The only other SGR established as a persistent hard X-ray source to date is SGR\,1806$-$20 \citep{mereghetti05int,molkov05}. For this source observations carried out with \xmm in the April 2003--October 2004 period, showed a progressive spectral hardening in the \mbox{1--10 keV} band, as the source increased its burst rate before the giant flare \citep{mte05}. The \igr observations displayed some evidence of a similar behaviour above \mbox{20 keV}. In fact its photon index varied from $\sim$1.9 in the period March 2003--April 2004 to $\sim$1.5 in September--October 2004 \citep{mereghetti05int}. A comparison of the hard X-ray luminosity of the two \sgr in the ``pre-flare'' state is subject to uncertainties in their distances. For \src a distance of 15 kpc has been derived based on its likely association with a young star cluster \citep{vrba00}, while for SGR\,1806$-$20 the distance is rather debated and has been variously estimated from 6.4 kpc to 15 kpc \citep{cameron05,mcclure05}. If we assume a distance of \mbox{15 kpc} for both sources we obtain similar \mbox{20--100 keV} luminosities: \mbox{$(1.5\pm0.3)\times10^{36}$ \lum} for \src and \mbox{$(1.2\pm0.1)\times10^{36}$ erg s$^{-1}$} for SGR\,1806$-$20.\\
\indent These results, together with the recent detections of
several AXPs in the hard X-ray range
\citep{molkov04,kuiper04,revnivtsev04,denhatog06,kuiper06}  with
\mbox{20--100 keV} luminosities similar or larger than those below
10 keV, indicate that non thermal magnetospheric phenomena are
energetically important in magnetars. Soft X-rays give only a
partial view  and broad band observations are required for a better
understanding of the physical processes occurring in these sources.
In this respect,  \src\,  being probably the first magnetar showing
evidence for variability in the hard X-ray range  and currently in a
moderately active state, is a good target to further explore
possible correlations between the persistent  emission and the
bursting activity. \indent \begin{acknowledgements}
\end{acknowledgements}
This work has been partially supported by the Italian Space Agency
and by the Italian Ministry for Education, University and Research
(grant PRIN-2004023189). The authors are grateful to the anonymous
referee whose valuable comments led to substantial improvements in
the paper.
\bibliographystyle{aa}
\bibliography{biblio}
\onecolumn
\begin{table}
\caption{Summary of the \sax observations of \src.}
\label{obs}
\centering
\begin{tabular}{ccccccc}
\hline\hline
Obs. & Date & MJD & LECS exposure & MECS exposure & PDS exposure & Period\\
\hline
A & 1997-05-12 & 50580 & 19.9 ks & 45.8 ks & 20.1 ks & $5.15719\pm0.00003$ s\\
B & 1998-09-15 & 51071 & 13.8 ks & 33.3 ks & 15.8 ks & $5.16026\pm0.00002$ s\\
C & 2000-03-30 & 51633 & 14.4 ks & 40.3 ks & 18.3 ks & $5.16709\pm0.00003$ s\\
D & 2000-04-25 & 51659 & 17.4 ks & 40.5 ks & 18.8 ks & $5.16765\pm0.00003$ s\\
E & 2001-04-18 & 52017 & 20.4 ks & 46.4 ks & 16.7 ks & $5.17277\pm0.00001$ s\\
F & 2001-04-29 & 52028 & 25.7 ks & 57.6 ks & 25.6 ks & $5.17298\pm0.00001$ s\\
G & 2001-11-05 & 52218 & --      & 1.3 ks  & 0.5 ks  & --\\
H & 2002-03-09 & 52342 & --      & --      & 47.6 ks & --\\
I & 2002-04-27 & 52391 & --      & 82.9 ks & --      & $5.18019\pm0.00002$ s\\
\hline
\end{tabular}
\end{table}
\begin{table}
\begin{minipage}[t]{\columnwidth}
\caption{Summary of the spectral results in the 0.8--10 keV energy
range. Errors are given at the 90\% confidence level.} \label{fits}
\centering
\begin{tabular}[c] {cccccccc}
\hline\hline
Obs. & Model & $N_{\rm H}$ & $\Gamma$ & $k_B T$ & $R_{bb}$\footnote{Radius at infinity assuming a distance of 15 kpc.} & Flux\footnote{Flux in the 2--10 keV range, corrected for the absorption.} & $\chi^{2}_{r}$ (d.o.f.) \\
& &  ($10^{22}$ $\rm cm^{-2}$) & & (keV) & (km) & ($10^{-11}$ erg cm$^{-2}$ s$^{-1}$) & \\
\hline
A & PL  & $1.4\pm0.2$ & $1.9\pm0.1$ & -- & -- & $0.92^{+0.04}_{-0.03}$ & 1.60 (98) \\
 & PL+BB  & $1.6^{+0.6}_{-0.4}$ & $0.9^{+0.3}_{-0.4}$ & $0.5\pm0.1$ & $4^{+2}_{-1}$ & $1.0\pm0.1$ & 1.12 (96) \\
 & PL+BB  & 2.6 (fixed) & $1.1^{+0.3}_{-0.2}$ & $0.44^{+0.03}_{-0.05}$ & $7^{+2}_{-1}$ & $1.11\pm0.06$ & 1.16 (97) \\
B & PL  & $2.4\pm0.2$ & $2.2\pm0.1$ & -- & -- & $2.6\pm0.1$ & 1.22 (113) \\
 & PL+BB  & $1.7^{+0.6}_{-0.5}$ & $1.5^{+0.5}_{-0.6}$ & $0.7\pm0.1$ & $3\pm1$ & $2.5\pm0.2$ & 1.16 (111) \\
 & PL+BB  & 2.6 (fixed) & $2.0\pm0.2$ & $0.5\pm0.1$ & $5\pm2$ & $2.7\pm0.2$ & 1.21 (112) \\
C & PL  & $2.0^{+0.4}_{-0.3}$ & $2.3\pm0.1$ & -- & -- & $0.95\pm0.04$ & 1.29 (84) \\
 & PL+BB  & $2\pm1$ & $1.7^{+0.3}_{-0.6}$ & $0.5\pm0.1$ & $5^{+7}_{-2}$ & $1.0^{+0.1}_{-0.2}$ & 1.09 (82) \\
 & PL+BB  & 2.6 (fixed) & $1.7\pm0.3$ & $0.45\pm0.05$ & $6^{+2}_{-1}$ & $1.1\pm0.1$ & 1.08 (83) \\
 D & PL  & $2.1\pm0.3$ & $2.4\pm0.1$ & -- & -- & $0.90\pm0.05$ & 1.06 (83) \\
 & PL+BB  & $2.2^{+0.9}_{-0.7}$ & $2.0^{+0.4}_{-0.5}$ & $0.5\pm0.1$ & $4^{+8}_{-2}$ & $0.9^{+0.1}_{-0.2}$ & 1.00 (81) \\
 & PL+BB  & 2.6 (fixed) & $2.1\pm0.3$ & $0.4\pm0.1$ & $7^{+4}_{-2}$ & $1.0\pm0.1$ & 0.99 (82) \\
 E & PL  & $3.6\pm0.2$ & $2.2\pm0.1$ & -- & -- & $3.5\pm0.1$ & 1.18 (121) \\
 & PL+BB  & $2.6^{+0.7}_{-0.9}$ & $1.8^{+0.3}_{-0.8}$ & $0.9^{+0.2}_{-0.1}$ & $1.6^{+0.6}_{-0.8}$ & $3.1^{+0.3}_{-0.2}$ & 1.13 (119) \\
 & PL+BB  & 2.6 (fixed) & $1.7^{+0.1}_{-0.2}$ & $0.9\pm0.1$ & $1.7^{+0.4}_{-0.5}$ & $3.1\pm0.2$ & 1.12 (120) \\
 F & PL  & $2.4^{+0.3}_{-0.2}$ & $2.3\pm0.1$ & -- & -- & $1.06^{+0.04}_{-0.03}$ & 1.35 (105) \\
 & PL+BB  & $2.4^{+0.8}_{-0.5}$ & $1.4\pm0.4$ & $0.5\pm0.1$ & $5^{+3}_{-1}$ & $1.1^{+0.1}_{-0.2}$ & 1.06 (103) \\
 & PL+BB  & 2.6 (fixed) & $1.5\pm0.3$ & $0.51^{+0.03}_{-0.05}$ & $5\pm1$ & $1.1\pm0.1$ & 1.05 (104) \\
 I & PL  & $1.9\pm0.3$ & $2.4\pm0.1$ & -- & -- & $0.62\pm0.03$ & 1.06 (94) \\
 & PL+BB  & $3\pm1$ & $2.2\pm0.3$ & $0.3\pm0.1$ & $8^{+25}_{-4}$ & $0.73^{+0.04}_{-0.09}$ & 0.94 (92) \\
 & PL+BB  & 2.6 (fixed) & $2.1\pm0.2$ & $0.37^{+0.04}_{-0.06}$ & $8^{+4}_{-2}$ & $0.69\pm0.04$ & 0.93 (93) \\
 \hline
\end{tabular}
\end{minipage}
\end{table}
\begin{table}
\begin{minipage}[t]{\columnwidth}
\caption{PDS count rates for the `off' and `on' source positions
during the observations of \src. Errors are given at 1\,$\sigma$.}
\label{pds} \centering
\begin{tabular}{ccccccccc}
\hline\hline
Region       & Energy band  & Obs. A & Obs. B & Obs. C & Obs. D & Obs. E & Obs. F & Obs. H\\
   & (keV)               & (cts\,/\,s) & (cts\,/\,s) & (cts\,/\,s) & (cts\,/\,s) & (cts\,/\,s) & (cts\,/\,s) & (cts\,/\,s) \\
\hline
OFF $-$     & 20--50   & $4.88\pm0.02$ & $4.67\pm0.02$ & $3.75\pm0.01$ & $3.69\pm0.01$ & $3.37\pm0.01$ & $3.38\pm0.01$ & $3.35\pm0.01$ \\
            & 50--150  & $6.87\pm0.02$ & $6.51\pm0.02$ & $5.23\pm0.02$ & $5.27\pm0.02$ & $4.75\pm0.01$ & $4.75\pm0.01$ & $4.73\pm0.01$ \\
OFF $+$     & 20--50   & $4.91\pm0.02$ & $4.67\pm0.02$ & $3.74\pm0.02$ & $3.71\pm0.01$ & $3.38\pm0.01$ & $3.40\pm0.01$ & $3.35\pm0.01$ \\
            & 50--150  & $6.83\pm0.02$ & $6.44\pm0.02$ & $5.24\pm0.02$ & $5.21\pm0.02$ & $4.76\pm0.01$ & $4.78\pm0.01$ & $4.75\pm0.01$ \\
ON\footnote{Background subtracted values.}                                                   & 20--50   & $0.27\pm0.03$ & $0.79\pm0.04$ & $0.11\pm0.03$ & $0.36\pm0.03$ & $0.45\pm0.03$ & $0.43\pm0.02$ & $9.33\pm0.02$ \\
            & 50--150  & $0.21\pm0.04$ & $0.12\pm0.04$ & $<0.03$ & $<0.02$ & $0.11\pm0.03$ & $0.06\pm0.03$ & $4.48\pm0.02$ \\
\hline
\end{tabular}
\end{minipage}
\end{table}
%
\begin{table}
\begin{minipage}[t]{\columnwidth}
\caption{Time resolved spectral results for observation E. The table gives the parameters of a blackbody component added to a fixed component with $N_{\rm H}=2.6\times 10^{22}$ $\rm cm^{-2}$, $\Gamma=2$, $k_B T=0.4$ keV, and $R_{bb}=7$ km (see  Section \ref{obsE} for details). Errors are given at 1\,$\sigma$.}
\label{fitE}
\centering
\begin{tabular}[c] {cccc}
\hline\hline
Time interval & $k_B T$ (keV) &  $R_{bb}$\footnote{Radius at infinity assuming a distance of 15 kpc.} (km) & $\chi^2_r$ (d.o.f.)\\
\hline
I   & $1.23^{+0.02}_{-0.03}$ & $1.8\pm0.1$    & 1.03 (69)\\
    & 1.1 (fixed)            & $2.17^{+0.03}_{-0.02}$ & 1.35 (70)\\
    & $1.29\pm0.01$ & 1.6 (fixed)            & 1.09 (70)\\
II  & $1.15^{+0.03}_{-0.02}$ & $1.6\pm0.1$    & 1.21 (74)\\
    & 1.1 (fixed)            & $1.77^{+0.02}_{-0.03}$ & 1.24 (75)\\
    & $1.16\pm0.01$ & 1.6 (fixed)            & 1.20 (75)\\
III & $1.16\pm0.04$ & $1.4\pm0.1$    & 0.98 (64)\\
    & 1.1 (fixed)            & $1.56^{+0.02}_{-0.03}$ & 1.01 (65)\\
    & $1.09\pm0.01$ & 1.6 (fixed)            & 1.04 (65)\\
IV  & $1.10\pm0.05$ & $1.5^{+0.2}_{-0.1}$    & 1.31 (45)\\
    & 1.1 (fixed)            & $1.50\pm0.03$ & 1.28 (46)\\
    & $1.07\pm0.01$ & 1.6 (fixed)            & 1.29 (46)\\
V   & $0.94\pm0.07$ & $1.5^{+0.3}_{-0.2}$    & 1.25 (29)\\
    & 1.1 (fixed)            & $1.09\pm0.04$ & 1.35 (30)\\
    & $0.92^{+0.01}_{-0.02}$ & 1.6 (fixed)            & 1.21 (30)\\
\hline
\end{tabular}
\end{minipage}
\end{table}

\end{document}